\documentclass[a4paper,11pt]{article}
    
    \usepackage{jcappub} 
    
    \usepackage[T1]{fontenc} 
    \usepackage[utf8]{inputenc}
    \usepackage{graphicx}	
    \usepackage{amsmath}	
    \usepackage{cleveref}
    \usepackage{multirow}
    \usepackage{tikz}
    \usepackage{comment}
    \usepackage{lipsum}
    \usepackage{bm}
    \usepackage{caption}
    \usepackage{float}

    \renewcommand{\vec}{\bm}

    \definecolor{lime}{HTML}{A6CE39}
    \newcommand{\orcidicon}{%
    	\begin{tikzpicture}
    	\draw[lime, fill=lime] (0,0) 
    		circle [radius=0.16] 
    		node[white] {{\fontfamily{qag}\selectfont \tiny ID}};
    	\draw[white, fill=white] (-0.0625,0.095) 
    		circle [radius=0.007];
    	\end{tikzpicture}
    	\hspace{-5mm}
    }
    \newcommand\orcidRDM{{\href{https://orcid.org/0000-0002-3842-9297}{\orcidicon}}}
    \newcommand\orcidIDM{{\href{https://orcid.org/0000-0001-5948-9689}{\orcidicon}}}
    
    \title{Unveiling the nature of SgrA* with the geodesic motion of S-stars}
    
    \author[a]{Riccardo Della Monica\orcidRDM{}}
    \author[a]{Ivan de Martino\orcidIDM{}}

    \affiliation[a]{Universidad de Salamanca, Departamento de Fisica Fundamental, P. de la Merced S/N, Salamanca, ES}
    
    \emailAdd{rdellamonica@usal.es}
    \emailAdd{ivan.demartino@usal.es}
    
    \abstract{Despite the huge improvements guaranteed by future GRAVITY observations of the S0-2 star, these will not be able to unveil the fundamental nature, whether black hole or wormhole, of the central supermassive object. Nevertheless, observing stars orbiting closer to the central gravitational source could allow to distinguish between the black hole and wormhole nature of this object at more than 5$\sigma$. 
    
    Firstly, we have used publicly available astrometric and spectroscopic measurements of the S0-2 star to constrain the metric around the supermassive object without finding any evidence either favouring or ruling out the wormhole nature. Secondly, we have designed a mock catalogue of future observations of the S0-2 star mirroring the accuracy and precision of GRAVITY. Afterwards, we firstly tested our methodology showing that our procedure recovers the input model, and subsequently we demonstrated that the constraining power of such a dataset is not enough to distinguish between black hole and wormhole. 
    Finally, we built some toy models  representing stars orbiting much closer the central object than S0-2. 
    We used these toy models to investigate which are the ideal orbital features and observational strategies to achieve our aim of unveiling the fundamental nature of the central supermassive object, demonstrating that a star with a period of the order of $\sim 5$ years and a pericentre distance of $\sim 5$ AU could identify the nature of the central object at almost 5$\sigma$ accuracy. }
    
\begin{document}
    
    \maketitle
    \flushbottom
    \section{Introduction}
    
    The centre of Milky Way has been subject of an intense observational program, firstly in the radio and subsequently at near-infrared and X-ray wavelengths, throughout almost the last thirty years leading to exhibit the existence of a point source supermassive object named  Sagittarius A* (Sgr A*)  \cite{melia2001, genzel2010}.  While stars orbiting around it are accelerated by gravity to speeds of 10.000 km/s, Sgr A* is moving at less than 1 km/s, according to accurate radio interferometry measurements. Hence, the monitoring of the stellar cluster orbiting around  Sgr A* \cite{eckart1997, ghez2000} sets up its mass to  $M\sim4\times10^6M_\odot$ concentrated in a region of only a few hundreds Schwarzschild radii (e.g., \cite{reid2004, eisenhauer,Bower2004}). These studies have been awarded with the Nobel Prize\footnote{\url{https://www.nobelprize.org/uploads/2020/10/advanced-physicsprize2020.pdf}}.
    
    Nevertheless, the intrinsic nature of this object is still a mystery. While the main interpretation of the observations leads to the conclusion that Sgr A* is a supermassive black hole (SMBH), intense studies have been carried out exploring alternatives and deepining our understanding of the Schwarzschild solution. As an example, studies of strong gravitational lensing due to a Schwarzschild black hole have been carried out and particularized to the SMBH in our Galactic Centre (GC) \cite{Kumar2000, Kumar2008}. Moreover, many studies have been carried out assuming that Sgr A* is a SMBH to investigate whether the underlying theory of gravity may differ from General Relativity (GR) or not. Those studies lead, for instance, to novel constraints on $f(R)$-gravity \cite{delaurentisdemartino, deMartino2021}, and Scalar-Vector-Tensor theory \cite{dellamonica2021orbital}.  Another intriguing hypothesis is made by placing at the GC  a dense concentration of self-gravitating fermionic dark matter (DM) \cite{rar2015}. This DM matter model that may provide a solution to the puzzling DM mystery and accommodate some small scale issues of the standard Cold Dark Matter paradigm \cite{deMartinoChakrabarty2020}, provides an interesting explanation to the peculiar radial accelerations of the G2 object \cite{becerra2020} and an interesting fitting to the orbital motion of S0-2 and other known stars \cite{becerra2021, arguelles2021}. Nevertheless, the SMBH nature of Sgr A* is far to be ruled out, and the model in \cite{rar2015} is strongly disfavoured when the radio and IR data of SgrA* are taken into account.
    
    Recently, several studies focused on recognizing the intrinsic nature of Sgr A*, whether a SMBH or a wormhole (WH) \cite{Guerrero2021, 2021arXiv211010002O,2021arXiv210608070J,Cheng2021, DeChang2019, Simonetti2020, Bambi2021}. For instance, observations of additional light rings in the shadow of such an object  surrounded by an accretion disk could be the signature of a black hole (BH) mimicker \cite{Guerrero2021, 2021arXiv211010002O}. Additionally, WHs models based on GR, brane-world gravity, and Einstein-Dirac-Maxwell theory have been constrained using the orbital motion of the S0-2 star \cite{2021arXiv210608070J}. Finally, 
    observations of weak  gravitational lensing mirroring the limitations of the near-infrared interferometer GRAVITY have been used to forecast the accuracy down to which deviations from a Schwarzschild BH can be detected \cite{Cheng2021}.
    
    With respect to previous analysis, we want to study a uniparametric class of models ranging from a Schwarzschild BH to a WH in light of the current observations of the S0-2 star, and in light of future observations made with the interferometer GRAVITY \cite{gravity2017}. Our aim is, on one side, to constrain the departure from a Schwarzschild BH, and on the other side to forecast the precision down to which GRAVITY will constrain such a departure near in future. In order to achieve our goals, first, we use publicly available data of the orbital motion of S0-2 star and estimate the departure from a Schwarzschild BH using a Monte Carlo Markov Chain (MCMC) algorithm. Second, we build mock catalogues mirroring the future observations of GRAVITY following the prescription given in \cite{grould}. Then, we firstly demonstrate that  our method is capable to recover our baseline model, and then we forecast the accuracy achievable with those observations. Finally, we
    build toy models to determine which type of observations we would need to detect a departure from a Schwarzschild BH using astrometric and interferometric measurements of a star's orbit. This manuscript is sectioned as follows: in Sect. \ref{sec:BB} we briefly summarize the theoretical model and its characteristics. In Sect. \ref{sec:s2_data}, we briefly summarize the data used in our analysis, how we build the orbital model and contrast it to the data using a MCMC algorithm, and we discuss our results. Then, in Sect. \ref{sec:s2_mock} we describe the  strategy used to create a mock catalogue for future observations of S0-2 star by GRAVITY, we test our pipelines to show their effectiveness to recover the baseline model, and finally we forecast the accuracy down to which we would be able in future to detect departure from Schwarzschild BH using S0-2 star.  In Sect. \ref{sec:S62_toy}, we study the case of other stars having a pericentre closer to the central object than S0-2. We build toy models and the corresponding mock catalogues to analyse under which observational conditions we could be able to exclude a departure from Schwarzschild BH. Finally, in Sect. \ref{sec:conclusions} we give our final discussions and expose our conclusions.

    \section{Black bounce model}\label{sec:BB}
    In GR, the Black Bounce (BB) metric refers to the one-parameter family of solutions, firstly introduced in \cite{simpson2019}, described by the following line element:
    \begin{equation}
        ds^2 = -\left(1-\frac{2M}{\sqrt{x^2+\alpha^2}}\right)dt^2+\left(1-\frac{2M}{\sqrt{x^2+\alpha^2}}\right)^{-1}dx^2+(x^2+\alpha^2)(d\theta^2+\sin^2\theta d\phi^2),
        \label{eq:metric}
    \end{equation}
    where we have set $G = c = 1$. This is a static and spherically symmetric family of solutions tuned by the parameter $\alpha$ \cite{simpson2019}. Indeed, setting $\alpha=0$ returns the Schwarzschild metric.
    The space-time coordinates $(t, x, \theta, \phi)$ employed in the metric element \eqref{eq:metric}, have the following domains
    \begin{align}
        &t\in(-\infty,+\infty),& &x\in(-\infty,+\infty),& &\theta\in(0,\pi),& &\phi\in(0,2\pi).&
    \end{align}
    
    Considering the line element $ds^2_{(2)}$ on hypersurfaces with $\{t= \textrm{const.}; x = \textrm{const.}\}$,  $(\theta,\phi)$ represent the usual spherical angular coordinates on a sphere whose surface area is $A = \int ds^2_{(2)} = 4\pi r^2$, where $r^2(x) = x^2 + \alpha^2$ can be regarded as the aerial radius. The parameter $\alpha$ defines the so-called \emph{throat} of the geometry, $r_{\rm th}^2 = \alpha^2$, representing an absolute minimum of the aerial radius
    (which is usually called \emph{bounce}, hence the name of this model). This feature allows to smoothly interpolate between a Schwarzschild BH  and a transversable WH. This can be seen by looking at the causal structure of the geometry:
    \begin{equation}
        g_{00} = 1-\frac{2M}{\sqrt{x^2+\alpha^2}} = 0 \quad \Rightarrow \quad x_H = \pm\sqrt{4M^2-\alpha^2}.
    \end{equation}
    Then, an horizon exists if and only if the following condition holds: $\alpha\leq 2M$; and, for each value of $\alpha$ in this range, the geometry has two horizons, one on each side of the throat. In particular:
    \begin{itemize}
        \item For $\alpha = 0$, then $x_H = \pm 2M$ and the geometry reduces to the usual Schwarzschild BH with mass $M$;
        \item For $0 < \alpha < 2M$, the metric possesses both an horizon and a non-null throat, but $r^2(x_H) = 4M^2 > r_{\rm th}^2$ and hence, the bounce is ``hidden'' behind the horizon. This solution corresponds to a regular BH (i.e. without a singularity) geometry \cite{Roman1983, Hayward2006};
        \item For $\alpha = 2M$, the horizon and the throat coincide, meaning that the geometry represents a WH connecting two universes. On the other hand, the bounce located at $x = 0$ is an extremal null throat and, hence, the WH is only one-way transversable \cite{Cano2019, simpson2019};
        \item Finally, for $\alpha > 2M$, the geometry is free of horizons and the bounce is totally exposed on both sides, meaning that the geometry represents a two-ways transversable WH \cite{Morris1988_1, Morris1988_2, Visser1989_1, Visser1989_2}.
    \end{itemize}

    The value $\alpha = 2M$ represents the separation between the BH/WH classes of geometries in this family.
    
    \subsection{Geodesic motion}
    Since our aim is to study the motion of massive test particles that undergo free-fall motion around a BB, we need to solve the geodesic equations related to the metric in Eq.  \eqref{eq:metric} for the time-like case. Such equations read:
   
    \begin{align}
        \ddot{t} = & - \frac{2 M x\dot{t}\dot{x}}{\Xi(x)\left(\alpha^{2} + x^{2}\right)^{3/2}},\label{eq:geodesic_t}\\
        \nonumber\\
        \ddot{x} = & \frac{Mx(\dot{x}^2-\dot{t}^2\Xi^2(x))}{\Xi(x)(\alpha^2+x^2)^{3/2}}-\Xi(x)(x\dot{\phi}^2\sin^2\theta+x\dot{\theta}^2)\label{eq:geodesic_x}\\
        \nonumber\\
        \ddot{\theta} = & - \frac{2 x \dot{x}\dot{\theta}}{\alpha^{2} + x^{2}} + \frac{\dot{\phi}^{2}\sin(2\theta)}{2},\label{eq:geodesic_theta}\\
        \nonumber\\
        \ddot{\phi} = & - \frac{2 x \dot{\phi} \dot{x}}{\alpha^2+x^2}- 2\dot{\phi} \dot{\theta}\cot(\theta).\label{eq:geodesic_phi}
    \end{align}
    where  we have defined $\Xi(x) \equiv 1-2M/\sqrt{\alpha^2+x^2}$. This system of second-order ordinary differential equations can be integrated numerically once initial conditions for the four coordinate functions $\{t(s_0), x(s_0), \theta(s_0), \phi(s_0)\}$ and their derivatives $\{\dot{t}(s_0), \dot{x}(s_0), \dot{\theta}(s_0), \dot{\phi}(s_0)\}$ are assigned at an initial point $s_0$ (that we take to be $s_0 = 0$ without loss of generality). In particular, since the metric is stationary, the particular value of $t(0)$ is unimportant for the following evolution of the geodesic (as can be seen from the fact the the geodesic equations \eqref{eq:geodesic_t}-\eqref{eq:geodesic_phi} are independent of the coordinate $t$), and for the particular case of a time-like geodesic describing the motion in space-time of a massive test particle, the normalization condition on the four-velocity yields $g_{\mu\nu}\dot{x}^\mu\dot{x}^\nu = -1$. From this condition, one relates one of the components of the four-velocity to the others. In particular, we normalize the four-velocity by expressing $\dot{t}(0)$ as a function of $\dot{x}(0)$, $\dot{\theta}(0)$ and $\dot{\phi}(0)$. This leaves us with 6 degrees of freedom for setting the initial conditions of the test particle, namely $\{ x(0), \theta(0), \phi(0), \dot{x}(0), \dot{\theta}(0), \dot{\phi}(0)\}$, which represent the spatial position and velocity at the initial time. When studying the orbital motion of celestial bodies, it is customary to recast this six initial conditions in terms of the Keplerian orbital elements of the orbit. These are: the time of passage at pericentre $t_p$, the semi-major axis of the orbit $a$, its eccentricity $e$, the orbital inclination $i$, the angle of the line of nodes $\Omega$ and the argument of the pericentre $\omega$. While in classical Newtonian gravity these parameters are constant and uniquely identify the orbit of a freely falling object in a central potential throughout the time, in GR they evolve with time \cite{poisson_will_2014}. This means that the orbital parameters assigned at the initial time only identify the osculating ellipse at that specific moment.

    \section{Data and Data Analysis of the S0-2 star}
    \label{sec:s2_data}
    
    The S0-2 star is a B-type star in the nuclear cluster orbiting the radio source SgrA* in the GC of our galaxy. Its orbit is characterized by an orbital period of $\sim 16$ years, a semi-major axis of  $\sim$ 970 AU and an high eccentricity of $\sim 0.88$ \cite{gillessen}. These orbital properties lead to a periastron distance from SgrA* of just $\sim$ 120 AU, which is as close as 1400 gravitational radii of the central compact object, and to a pericentre orbital speed of $\sim 7700$ km/s (which is $\sim2.5\%c$), making the pericentre passage of S0-2 a great opportunity to study relativistic effects \cite{gravity2018, do, gravity}, as well as to probe the underlying theory of gravity \cite{deMartino2021, dellamonica2021orbital}. Indeed, the sky-trajectory and radial velocity of the S0-2 star have been the subject of continuous monitoring during the last three decades, through astrometric and spectroscopic observations, culminating in 2018 with the last pericentre passage. Observations performed at pericentre revealed for the first time a departure from the previously recorded Keplerian motion of the star caused by general relativistic effects in the observed orbit \cite{gravity2018, do, gravity} and opening a new avenue to test gravity \cite{deMartino2021, dellamonica2021orbital, borka2021, benisty2021dark}. 
    
   Here, we use publicly available data for the S0-2 star to probe the nature of the central object. In particular, by fitting the orbit derived from the geodesic motion in Eqs. \eqref{eq:geodesic_t}-\eqref{eq:geodesic_phi} to the astrometric data, our goal is to constrain the parameter $\alpha$ in the BB metric that allows to discern between a BH and a WH nature of the gravitational source.
    
    \subsection{Data}
    \label{sec:data}
    Let us start to briefly summarize the dataset. For a more comprehensive description we refer to Supplementary material of \cite{deMartino2021}.
    The dataset that we used for our analysis is composed of three distinct kinds of measurements:
    
    \begin{itemize}
    	\item \textbf{Astrometric positions:} corresponding to the sky-projected position of the S0-2 star, as observed in the near-infrared band. In particular, we use the 145 data points that are publicly available in the electronic version of \cite{gillessen}. These data provide with right ascension (R.A.) and declination (Dec) for S0-2 relative to the  ‘GC infrared reference system’ \cite{Plewa} and, hence relative to the point on the sky where SgrA* is expected to be\footnote{Except for an offset and drift of SgrA* with respect to the origin of the reference frame that must taken into account when fitting an orbital model on the astrometric data \cite{Gillessen_2009a, gillessen, Plewa} and that constitutes a limiting factor in the precision of the measurement of relativistic effects on the orbit \cite{gravity}.}. These data cover the periods:
    	from 1992.224 to 2002 with the observations
    	made at the ESO New Technology Telescope  \cite[NTT, ][]{hoffman}, by means of the speckle camera SHARP, which provided with astrometric accuracy of order 3.8 mas; from 2002 with observations performed at the Very Large Telescope (VLT) using the adaptive optics assisted infrared camera NAOS+CONICA (NACO) instrument \citep{lenzen, rousset}, which provided with an astrometric accuracy of $\approx 400 \mu$as. 
    	\item \textbf{Radial velocities (RV):}  the measured line-of-sight velocity of S0-2 whose sign is assumed to be positive during the approaching phase and negative when the star recedes. A total of 44 data points
    	come from Brackett-$\gamma$ line shift spectroscopic measurements. In particular, up to 2003 these measurements have been done with NIRC2, a near-infrared Adaptive Optic imager and spectrometer located at the Keck observatory \citep{Ghez_2003}; later recordings of the radial velocity were made using SINFONI, the Spectrograph for INtegral Field Observations in the Near Infrared, mounted at the VLT \citep{eisenhauer2003, bonnet}.
    	\item \textbf{Orbital precession:} in 2020, the Gravity Collaboration was able to measure the rate of periastron advance for the S0-2 star \cite{gravity} by using accurate astrometry from the GRAVITY interferometer at VLT, operating in the K band. Since the astrometric dataset for the S0-2 star from the GRAVITY instrument is not publicly available, we only use their measurement of the orbital precession. In particular, they measured a putative parameter $f_{\rm SP}$ that multiplies the general relativistic orbital precession for a massive test particle around an object with mass $M$:
    		\begin{equation}
    			\Delta\omega = \Delta\omega_{\rm GR}f_{\rm SP} =  \frac{6\pi GM}{a^2c^2(1-e^2)}f_{\rm SP}
    			\label{eq:gr_precession}
    		\end{equation}
    		where $a$ is the star's semi-major axis, $e$ its eccentricity and $\omega$ the argument of pericentre. In particular,  $f_{\rm SP} = 0$ would be consistent with a closed elliptical orbit of Newtonian gravity, while $f_{\rm SP} = 1$ would imply a
    		precessing orbit as predicted by GR.
    		The measured value is $f_{\rm sp}=1.10\pm 0.19$ which excludes Newtonian gravity at 5$\sigma$ and is totally compatible with GR. Since the space-time geometry for a BB/WH is different from that of a Schwarzschild BH (for which Eq. \eqref{eq:gr_precession} holds),
    		we expect that the measurement of the orbital precession can help to constrain the BB metric in Eq. \eqref{eq:metric}.
    		\end{itemize}

    \subsection{Orbital model and observational effects}
    \label{sec:orb_model}
    We compute numerically the fully relativistic sky-projected mock trajectory for S0-2  given its orbital elements (for a more details we refer to  the Supplementary Materials of \cite{deMartino2021}). Here, we report the main steps of the algorithm and the main physical and observational effects that must be taken into account to produce an accurate mock orbit.
    
    First, a unique metric for the space-time geometry must be fixed, by assigning values to the free parameters in Eq. \eqref{eq:metric}. Namely, the mass $M$ of the central object and the parameter $\alpha$ defining its fundamental nature. Geodesic equations are formulated on top of this metric and initial data for the position of the star are assigned by converting the Keplerian elements $(t_p, a, e, i, \Omega, \omega)$ into the initial position and velocity in the reference frame of the gravitational source. An additional orbital parameter, namely the orbital period $T$, is provided to the algorithm which serves to compute the mass $M$ of the central object (which is hence not an independent parameter) through Kepler's third law, $M := 4\pi^2a^3/GT^2$ (see also \cite{grould, gillessen}). The geodesic equations are then integrated numerically via an adaptive-step-size 4(5) Runge-Kutta integrator, both forward and backward in time (in order to cover the time span for which observations are available), starting from the initial time, that is taken to be the last apocentre passage, $t_a = t_p-T/2\sim2010.35$. The integrated orbit, namely a parametric array of coordinates as a function of proper time $\{t(s), x(s), \theta(s), \phi(s)\}$, must be converted into the physical quantities that one observes experimentally: the relative R.A. and Dec of the star, its line of sight apparent velocity, and the pericentre advance.
    
    The astrometric positions can be obtained via a geometric projection of the space-time coordinates, through the Thiele-Innes elements \cite{thieleinnes} and applying the classical Rømer delay formula in order to account for the modulation of the time of arrival of the light emitted by the star when is farther away or closer to Earth during its orbital motion. Converting from spatial positions of the star to angular separations (namely,  R.A. and Dec) requires the knowledge of the distance $R$ of SgrA* from Earth, which we leave as a free parameter of our model. An additional observational effect on the astrometric position that cannot be neglected \cite{Gillessen_2009a} is a zero-point offset and drift of central mass with respect to the reference frame resulting in five additional parameters $(x^*, y^*, v_x^*, v_y^*, v_z^*)$ to be fitted. Specifically, those are the position ($x^*, y^*$) of SgrA* on the sky-plane, its drift velocity ($v_x^*, v_y^*$), and an offset in the radial velocity $v_z^*$ when correcting spectroscopic observations to the Local Standard of Rest (LSR).
    
    Radial velocities are measured via the detected frequency shift in spectroscopic measurements of the light emitted by the star,
    \begin{equation}
        z:=\frac{V_R}{c} = \frac{\Delta\nu}{\nu}.
    \end{equation}
    While the kinematic line-of-sight velocity of the star accounts for much of this shift, through the Doppler formula, other effects produce significant frequency shift and must be taken into account in order to properly determine the observed radial velocity of the star. First of all, since the S0-2 star at pericentre reaches an orbital velocity that is $\sim 2.5\%$ of the speed of light in vacuum, the special relativistic time dilation cannot be neglected and produces an additional frequency redshift that is accounted for in the special relativistic Doppler formula:
    \begin{align}
    	z_{SR} + 1 = \frac{\sqrt{1-\frac{v^2}{c^2}}}{1-v_{\rm los}/c},
    \end{align}
    where $v$ is the norm of the spatial velocity of the star and $v_{\rm los}$ is its projections along the line of sight. Furthermore, since the star gets as close as $\sim 1400$ gravitational radii to SgrA*, the strong gravitational field generated by the source at such proximity produces an additional gravitational time dilation, resulting in an extra general-relativistic redshift component, given by
    \begin{equation}
        z_{GR} + 1 = \frac{1}{\sqrt{|g_{00}(t,\vec{x})|}}.
    \end{equation}
    The joint contribution of this two effects (whose frequency shifts must be multiplied between each other) yields an additional $\approx 200$ km/s in the apparent radial velocity of the star around pericentre that has been observed by the two independent groups in 2018 \cite{gravity2018, do}.
    
\begin{figure}[t]
        \centering
        \includegraphics[width = \textwidth]{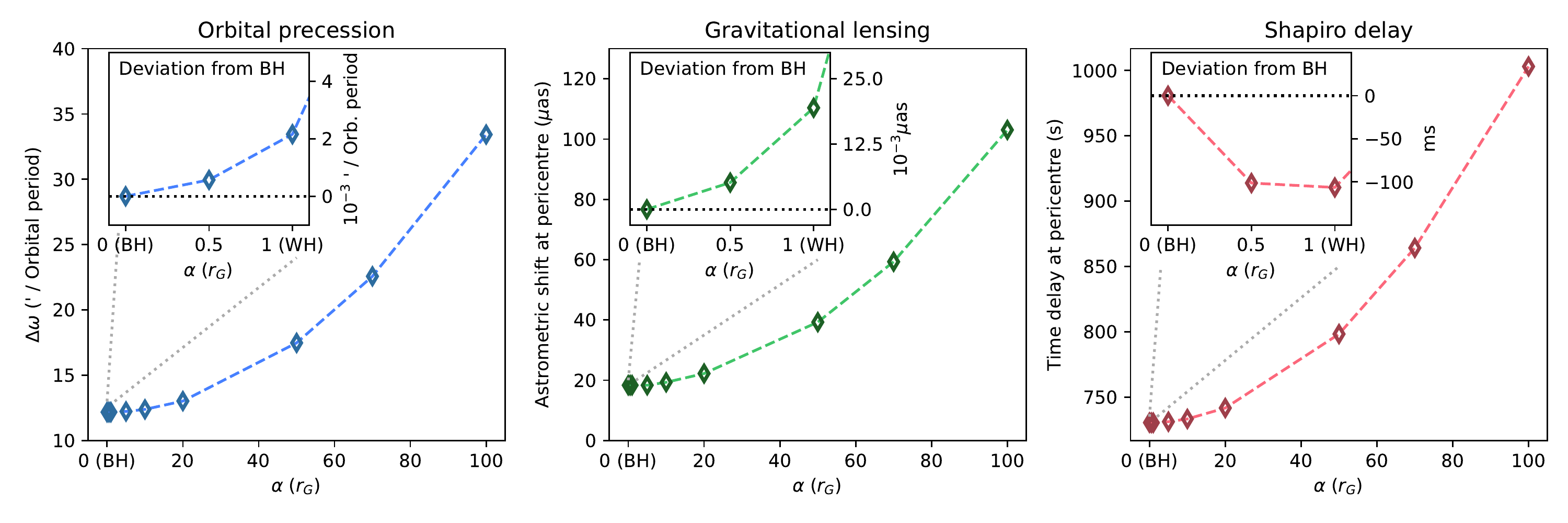}
        \caption{The impact of different relativistic effects on the motion of the S0-2 star, as a function of $\alpha$ in units of gravitational radii of the central source. In particular: \emph{(left panel)} the periastron advance of the star on its orbital plane in units of arcminutes per orbital period; \emph{(central panel)} the astrometric impact in units of $\mu$as of the gravitational lensing on the observed position of the star at pericentre (when this effect is maximized); \emph{(right panel)} the value of the Shapiro delay in seconds as measured when the star is at pericentre. The inset plots show the deviation of such effects from the $\alpha = 0$ case, in the interval $\alpha \in[0,1]$.}
        \label{fig:rel_effects}
    \end{figure}

    In Figure \ref{fig:rel_effects} we show the impact of different relativistic effects on the motion of the S0-2 star as a function of the BB parameter $\alpha$ in Eq. \eqref{eq:metric}. In particular, we report from the left to the right panel the rate of orbital precession, the astrometric impact of the gravitational lensing, and the time delay respectively. More specifically, in each panel we report the total magnitude of each effect, while in the insets we report the relative difference between the corresponding effect in the BB metric and the one in the standard  Schwarzschild metric. In details, we show:
    
    \begin{itemize}
        \item The rate of orbital precession $\Delta\omega$ of the star. This effect can be directly estimated from the integrated orbit in space-time coordinates. More specifically, we can look at the $(\phi,\dot{r})$ relation. The radial velocity $\dot{r}$ of the star vanishes at radial turning points on the orbit, namely the apocentre and the pericentre. Thus, the difference between the values of the $\phi$ coordinate corresponding to two successive apocentre passages gives the angle $\Delta\phi$ spanned by the star during one orbital period, from which one can get the periastron advance $\Delta \omega = \Delta\phi - 2\pi$. For the BH case, $\alpha = 0$, this effects yields $\sim 12$ arcminutes per orbital period and it increases with $\alpha$, gaining a factor $\sim 3$ when $\alpha = 100\,r_g$.
        \item The astrometric impact of the gravitational lensing (GL) on the observed position of the star at pericentre, i.e. when this effect is maximized. Following \cite{grould}, we reconstruct the lensed position of the star by performing a ray-tracing algorithm from the observer position (at distance $R$ from the source). In particular, we build a $5\times 5$ grid of (RA, Dec) positions on the image-plane of the observer centered on the unlensed position of the star (obtained by a simple geometric projection with the Thiele-Innes constants), with an appropriate grid step-size; for each position on the grid we integrate a null-geodesic backward in time, corresponding to a photon reaching the observer from that particular direction; for each integrated null-geodesic we identify the one that reaches the minimum euclidean distance from the star along its path; we build a new grid centered on such minimum-distance direction with a step-size that is one-forth of the previous one (so to cover an entire grid-cell) and we iterate this process until the photon reaches a minimum distance from the star of $\sim 10^{-2}\;r_g$. From the comparison between the unlensed and lensed positions, we finally compute the astrometric shift reported in the central panel of Figure \ref{fig:rel_effects}. As we can see, this effect yields a maximum shift of $\sim$20 $\mu$as at pericentre for the BH case, $\alpha = 0$, which goes up to $\sim$ 100 $\mu$as for $\alpha = 100$ $r_g$. As we show in the corresponding inset plot, the astrometric shift due to GL differs only on the order of $\sim 10^{-2}\;\mu$as between the BH and the WH case, i.e. $\alpha = 1$. Therefore,  we  neglect this effect in our analysis.
        \item The time delay, also known as Shapiro delay, due to the passage of the photon in the gravitational field of the central source, during its propagation from the star to the observer. This effect was estimated by comparing the proper time of the GL-computed photon from the previous point and the coordinate time of the observer and yields a value of $\sim\,730$ s (the corresponding estimated astrometric shift due to the motion of the star during this lapse of time is of units of $\mu$as) at pericentre for the case $\alpha = 0$. While the WH case, $\alpha = 1$, differs from the previous one by only $\sim -100$ ms. Such a small difference in time gives very little astrometric shifts, on the order of a few nano arcsec, between the two limiting cases of interest and therefore we neglect this contribution in our analysis.
    \end{itemize}
    
    \subsection{Monte Carlo Markov Chain}

    We explore a 14-dimensional parameter space, $\bar{\vec{\theta}} =$ ($T$, $t_p$, $a$, $e$, $i$, $\Omega$, $\omega$, $R$, $x^*$, $y^*$, $v_x^*$, $v_y^*$, $v_z^*$, $\alpha$), employing a MCMC algorithm, specifically we rely on \texttt{emcee} \cite{emcee}, to fit the mock orbits produced as discussed in section \ref{sec:orb_model} to the publicly available data presented in section \ref{sec:data}. In particular, extracting a random sample  from a previously assigned prior distribution $\Pi(\vec{\theta})$, we build the mock astrometric positions, radial velocities  and orbital precession 
    corresponding to that specific set of parameters and we estimate the posterior probability of $\bar{\vec{\theta}}$ as:
    \begin{align}
    	\log P(\bar{\vec{\theta}}|\textrm{ data}) =& -\frac{1}{2}\sum_i\biggl(\frac{\textrm{R.A.}_i(\bar{\vec{\theta}})-\textrm{R.A.}_{{\rm obs}, i}}{\sqrt{2}\sigma_{\rm R.A.}}\biggr)^2-\frac{1}{2}\sum_i\biggl(\frac{\textrm{Dec}_i(\bar{\vec{\theta}})-\textrm{Dec}_{{\rm obs}, i}}{\sqrt{2}\sigma_{\rm Dec}}\biggr)^2+\nonumber\\
    	& -\frac{1}{2}\sum_i\biggl(\frac{\textrm{RV}_i(\bar{\vec{\theta}})-\textrm{RV}_{{\rm obs}, i}}{\sqrt{2}\sigma_{\rm RV}}\biggr)^2-\frac{1}{2}\biggl(\frac{\Delta\omega(\bar{\vec{\theta}})/\Delta\omega_{\rm GR}-f_{{\rm SP}}}{\sqrt{2}\sigma_{f_{\rm SP}}}\biggr)^2+\nonumber\\
    	& + \Pi(\bar{\vec{\theta}})
    	\label{eq:likelihood}
    \end{align}
    
    The factor $\sqrt{2}$ multiplying the uncertainties on all the measurements
    is due to the fact that the precession estimate encoded in the parameter $f_{\rm SP}$ is not a direct measurement, but it has been done by using the same astrometric and spectroscopic dataset that we use (plus the data for the star's position and velocity at pericentre, measured with the GRAVITY interferometer, which are not publicly available). Thus, to be conservative, we follow \cite{deMartino2021, dellamonica2021orbital} and assume that all the data appear twice in the likelihood and thus we divide by 2 all the terms in the likelihood in order not to double-count data points. For this particular analysis, we used uniform priors an all the parameters of our model, except for the subset of parameters ($x^*$, $y^*$, $v_x^*$, $v_y^*$, $v_z^*$) for which we have information coming from independent sets of data and analysis, which allowed us to place Gaussian priors on these parameters. In particular, from \cite{Plewa} we can place limits on the sky-projected position of SgrA* in the radio-reference frame, while for the prior on the LSR offset, we expect the motion of SgrA* to be below 1 km/s \cite{gillessen} and thus use a Gaussian prior centered on $v^*_z = 0$ km/s with an uncertainty given by that of the LSR of $\sim 5$ km/s. The complete set of priors used in our MCMC analysis is reported in Table \ref{tab:priors}. Finally, the
    convergence of our MCMC analysis is assessed by means of the integrated autocorrelation time of the walkers.

    \begin{table}[]
    \centering
    \setlength{\tabcolsep}{20pt}
    \renewcommand{\arraystretch}{1.2}
    \begin{tabular}{@{}lcccc@{}}
    \hline\hline
    \multicolumn{2}{c}{}          &  \multicolumn{2}{c}{Uniform priors}     \\[0.5em]\cline{3-4}
    Parameter & Units & Start             & End     & Best fits             \\ \hline
    $T$   & yr                      & 15.5           & 16.5    &        $16.032\pm0.029$             \\
    $t_p$ & yr                        & 2018.0        & 2018.5  &       $2018.363\pm0.024$                    \\
    $a$ & as                          & 0.115           & 0.135  &      $0.1255\pm0.0011$                     \\
    $e$  &       -                  & 0.88          & 0.90         &     $0.8832\pm0.0024$            \\
    $i$ & degree               & 130           & 140            &    $134.07^{+0.50}_{-0.51}$            \\
    $\omega$ & degree                      & 65            & 70   &   $65.45\pm0.75$                 \\
    $\Omega$ & degree                      & 225           & 230   & $227.07^{+0.80}_{-0.79}$               \\
    $R$  & kpc                     & 8.0            & 8.5          &  $8.11\pm0.22$             \\
    $\alpha$                      &      $r_g$             &      0           & 100 & $\lesssim 42.59\, (1\sigma)$      \\[1em] \hline
    \multicolumn{2}{c}{}          &  \multicolumn{2}{c}{Gaussian priors} & Best fits     \\[0.5em]\cline{3-4}
    Parameter & Units & $\mu$             & $\sigma$                  \\ \hline
    $x^*$ & mas                   & -0.2             & 0.2          &     $-0.30\pm0.16$                \\
    $y^*$ & mas                   & 0.1              & 0.2          &       $0.08\pm0.19$              \\
    $v_{x}^*$ & mas/yr        & 0.02               & 0.2            &       $0.077^{+0.053}_{-0.054}$               \\ 
    $v_{y}^*$ & mas/yr       & 0.06              & 0.1              &   $0.132\pm0.064$              \\
    $v_{z}^*$ &  km/s       & 0              & 5                    &   $-2.8^{+4.4}_{-4.5}$        \\ \hline\hline
    \end{tabular}
    \caption{The set of priors used for our analysis on S0-2 star. Gaussian priors are taken from \cite{Plewa, gillessen}.}
    \label{tab:priors}
    \end{table}
    
    \subsection{Results and discussions}
    
\begin{figure}
    	\includegraphics[width = \textwidth]{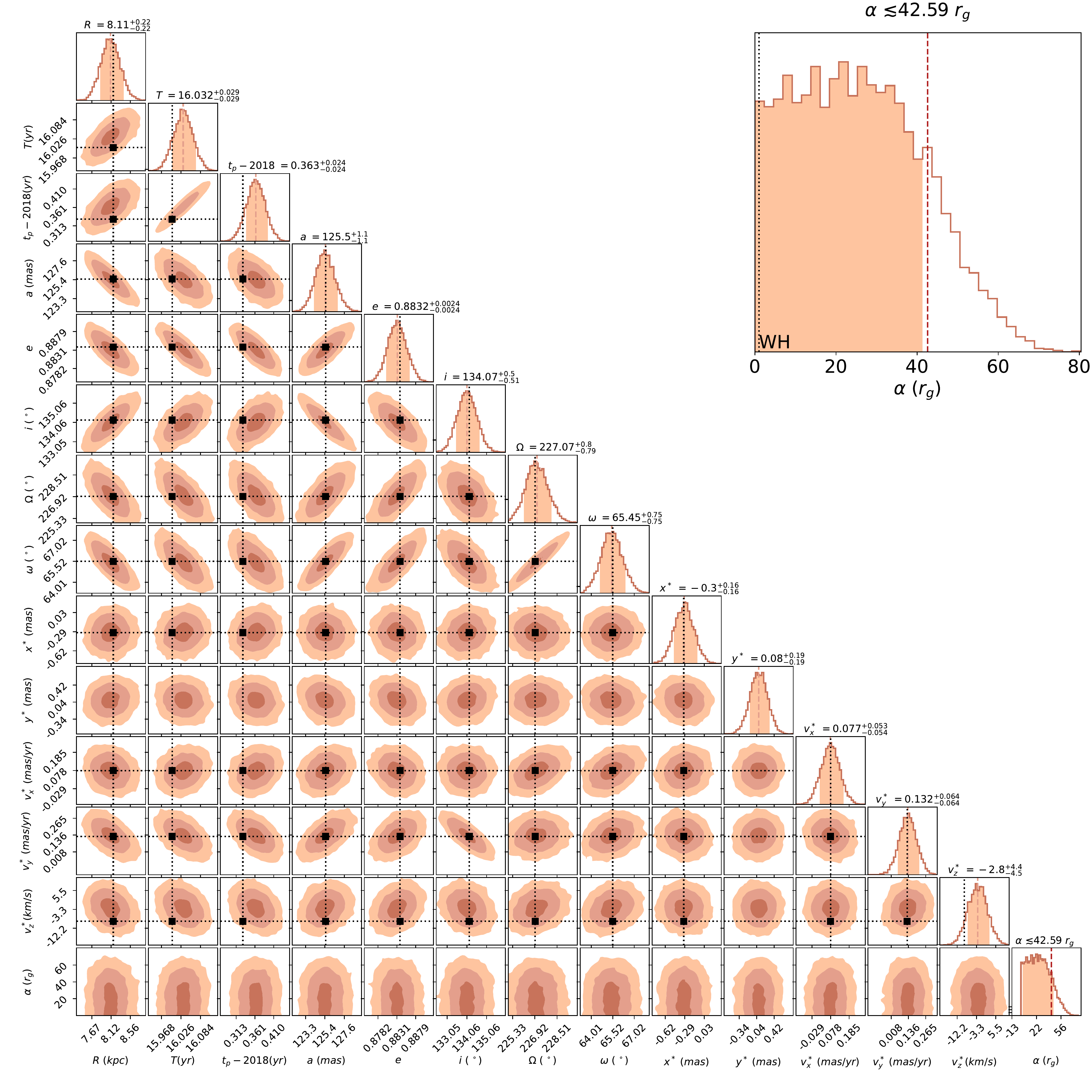}
    	\caption{The posterior distribution of the 14-dimensional parameter space of our model. Shaded regions on the contour plots show the 1$\sigma$, 2$\sigma$ and 3$\sigma$ contours respectively, while the shaded intervals in the histograms report the $1\sigma$ credible interval for the parameters (which is also reported numerically above). Black squares report the best fitting values from \cite{gillessen}. The inset plot zooms onto the posterior distribution of the $\alpha$ parameter. As can be seen, at the present stage of precision, we are not able to distinguish between a BH and and WH (the limit is marked by a dotted vertical black line), using observational data.}
    	\label{fig:posterior_with_data}
    \end{figure}

    We report in Figure \ref{fig:posterior_with_data} a corner plot showing the full posterior distribution of our 14-dimensional parameter space. From darker to lighter colours we represents the 68\%, 95\%, and 99\% of the confidence levels. On the top of each column, we report the best fitting value and 1$\sigma$ error of each parameter. On each 2D contour plot, we also report, as a black square, the  best fitting value of the corresponding parameters as obtained in \cite{gillessen}. Clearly, our estimates of the parameters are  compatible within 1$\sigma$  with previous ones  \cite{gillessen}. Moreover, we report  an inset plot of the posterior distribution of our parameter of interest ({\em i.e.} $\alpha$) in units of the gravitational radius $r_g = 2GM/c^2$. It is worth to note that, using current publicly available data for the S0-2 star, we are not able to distinguish between a BH and a WH (both natures fall well within our credible interval for the parameter $\alpha$). Nevertheless, it is remarkable that we  are able to exclude two-way transversable WHs whose throat is characterized by a radius $\alpha \gtrsim 70 r_g$, at 3$\sigma$ level.
    
    \section{Recognising the nature of Sgr A$^*$ from forthcoming observations of the S0-2 star}
    \label{sec:s2_mock}
    
    In the previous section we have demonstrated the current inability to distinguish between the BH nature and the WH one for SgrA* with the current available dataset on the motion of the S0-2 star.
    This is rather expected since the astrometric accuracy of such observations is quite poor (compared to  the accuracy provided by the GRAVITY interferometer that can be as precise as $\sim$ 10 $\mu$as) and, evidently, not able to reveal the slight change in the geodesic motion of the star that results from the different natures of Sgr A*. On the other hand, our results in Figure \ref{fig:posterior_with_data} show that, although poor in terms of discerning the two natures, we do succeed in placing an upper limit on the parameter $\alpha$ and we expect this upper limit to improve by a fair amount if we could be able to repeat the same analysis with more precise data. Observations of an entire orbit of S0-2 from the GRAVITY instrument, for example, not only would provide with much smaller astrometric error bars, but would also come with the great advantage of not having to introduce the offset and drift parameters ($x^*$, $y^*$, $v_x^*$, $v_y^*$, $v_z^*$) that are required for the NACO observations \cite{gravity2018, gravity}. 
    Thus, we would have better data and a lower dimensionality of the parameter space to constrain. 
    Nevertheless, astrometric data from the Gravity Collaboration are not yet publicly available, and they would only cover a small fraction of an entire orbital period, anyway. This motivates us to proceed further with our analysis and generate a mock catalogue of observations for the S0-2 star mirroring  measurements of GRAVITY instrument (starting in 2018 with the pericentre passage) that will follow the orbit of the star for entire orbital periods,  mimicking the observational strategy of the Gravity Collaboration and assuming as baseline model a 
    Schwartzschild BH, {\emph{i.e.}, $\alpha=0$}.
    In section \ref{sec:mock_catalogue} we report the details of the algorithm that we employ to produce such mock observations. In section \ref{sec:testing_methodology} we test the effectiveness of our pipelines to recover the input parameters.
    Finally, we repeat our complete MCMC analysis on the mock catalogue (leaving $\alpha$ as a free parameter) to forecast the precision down to which we could be able to
    place constraints on the nature of SgrA* with forthcoming dataset.
    The results of this analysis are reported in section \ref{sec:mock_results}.
    
    \subsection{Mock Catalogue}
    \label{sec:mock_catalogue}
    
    In this section we present our mock catalogue mirroring the observational limitations of the VLT instruments GRAVITY (for the astrometry) and SINFONI (for the spectroscopic measurements leading to radial velocity estimates). We build the orbital model for the S0-2 star as presented in section \ref{sec:orb_model}. In our baseline model, we assume: \emph{(i)} the Schwartzschild BH geometry (i.e., $\alpha = 0$),  and \emph{(ii)} we set our  \emph{true} parameters to their best fitting values from  GRAVITY \cite{gravity}. These values are reported in Table \ref{tab:s2_gravity}. Finally, \emph{(iii)} we  prescribe that the GRAVITY instruments performs observations as follows\footnote{We assume, as already happened for the pericentre passage in 2018, that the sampling of the observations is more dense around pericenter and more rarefied elsewhere  \cite{gravity}.}  (for more details we refer to \cite{gravity}):
    \begin{table}
    \centering
    \renewcommand{\arraystretch}{1.5}
    \setlength{\tabcolsep}{10pt}
    \resizebox{\textwidth}{!}{%
    \begin{tabular}{l|cccccccr}
    \hline
     Parameter & $R$ & $a$ & $e$ & $i$ & $\Omega$ & $\omega$ & $T$ & $t_p$     \\ 
     & kpc & mas & & ($^\circ$) & ($^\circ$) & ($^\circ$) & yr & yr \\ \hline
    Value & 8.2467 & 125.058 & 0.884649 & 134.567 & 228.171 & 66.263 & 16.0455 & 2018.37900 \\
    Error & 0.0093 & 0.041 & 0.000066 & 0.033 & 0.031 & 0.031 & 0.0013 & 0.00016\\ \hline
    \end{tabular}}
    \caption{The distance of SgrA* and the orbital parameters of the S0-2 star as measured by the Gravity Collaboration \cite{gravity}.}
    \label{tab:s2_gravity}
    \end{table}

\begin{figure}
    	\includegraphics[width = \textwidth]{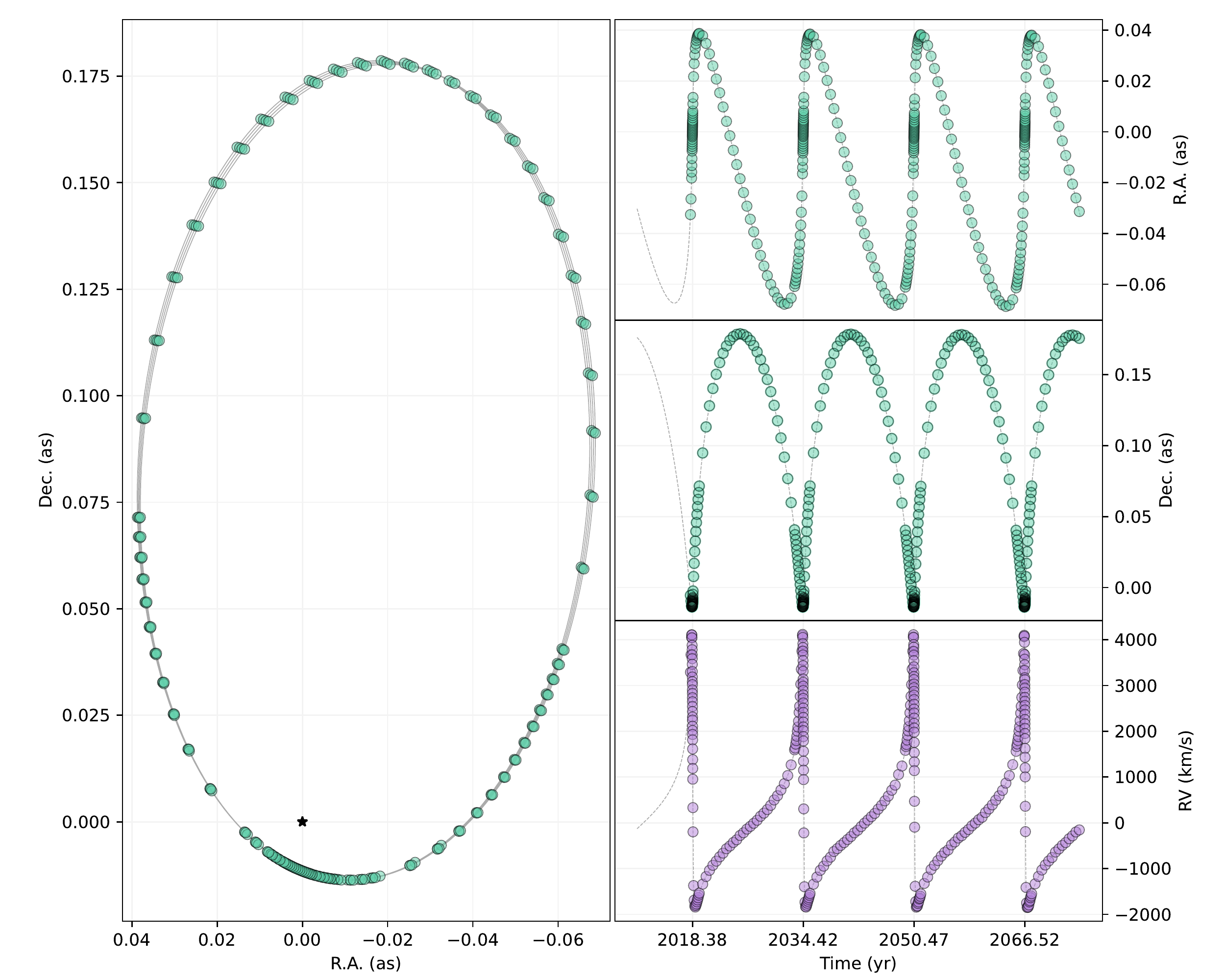}
    	\caption{The figure show our mock observations: astrometric positions (left panel and two top panels on the right) and radial velocity (right bottom panel) for the S0-2 star. }
    	\label{fig:mock_cataloge_s2}
    \end{figure}

    \begin{itemize}
    	\item one observation per day in the two weeks centered on the pericentre passage;
    	\item one observation every two nights in a month centered on the pericentre passage;
    	\item one observation per week in the two months centered on the pericentre
    	\item one observation per month in the rest of the year of pericentre passage;
    	\item two observations per year in the rest of the years.
    \end{itemize}
    We thus build an array of observation times accordingly and use it to compute astrometric positions and radial velocity measurements at those times. This comes under the additional assumption that spectroscopic and astrometric measurements are done on the same nights. Our geodesic integration is carried over for 2 entire periods of the S0-2 star.
    Moreover, we assume that the instruments are always operated in ideal conditions, acquiring data with an accuracy set to their maximal nominal values: $\sigma_A = 10\;\mu$as for the GRAVITY astrometry and $\sigma_{\rm RV} = 10$ km/s for the SINFONI radial velocities. Such uncertainties are taken into account by adding to the values given out by the orbital model a Gaussian noise with zero-mean  and FWHM given by $\sigma_{\rm A}$ and $\sigma_{\rm RV}$ for the astrometric data and radial velocities, respectively. Finally, as already mentioned, we assume that the GRAVITY observations always provide astrometry relative to SgrA*, meaning that we assume the central object to be fixed in the origin of the celestial reference frame, allowing us to avoid the introduction of offset and drift parameters of the reference frame. The resulting mock catalogue is depicted in Figure \ref{fig:mock_cataloge_s2}.

    \subsection{Testing the catalogue and the methodology}
    \label{sec:testing_methodology}
    
    In order to test the validity of our mock catalogue and of our fitting methodology, we applied our MCMC algorithm to the mock catalogue for the S0-2 star by fixing $\alpha = 0$ in the metric and letting the other parameters be fitted.
    The likelihood that we employed for this analysis is basically the same as in Eq. \eqref{eq:likelihood}, but removing the terms corresponding to the orbital precession (which is now naturally accounted for in the astrometric mock positions of the star) and, thus, removing the $\sqrt{2}$ factors at the denominators. We rewrite it here for sake of clearness:
    
    \begin{align}
    	\log P(\bar{\vec{\theta}}|\textrm{ data}) =& -\frac{1}{2}\sum_i\biggl(\frac{\textrm{R.A.}_i(\bar{\vec{\theta}})-\textrm{R.A.}_{{\rm obs}, i}}{\sigma_{\rm R.A.}}\biggr)^2-\frac{1}{2}\sum_i\biggl(\frac{\textrm{Dec}_i(\bar{\vec{\theta}})-\textrm{Dec}_{{\rm obs}, i}}{\sigma_{\rm Dec}}\biggr)^2+\nonumber\\
    	& -\frac{1}{2}\sum_i\biggl(\frac{\textrm{RV}_i(\bar{\vec{\theta}})-\textrm{RV}_{{\rm obs}, i}}{\sigma_{\rm RV}}\biggr)^2 + \Pi(\bar{\vec{\theta}}).
    	\label{eq:likelihood2}
    \end{align}

\begin{figure}[htbp]
    	\includegraphics[width = \textwidth]{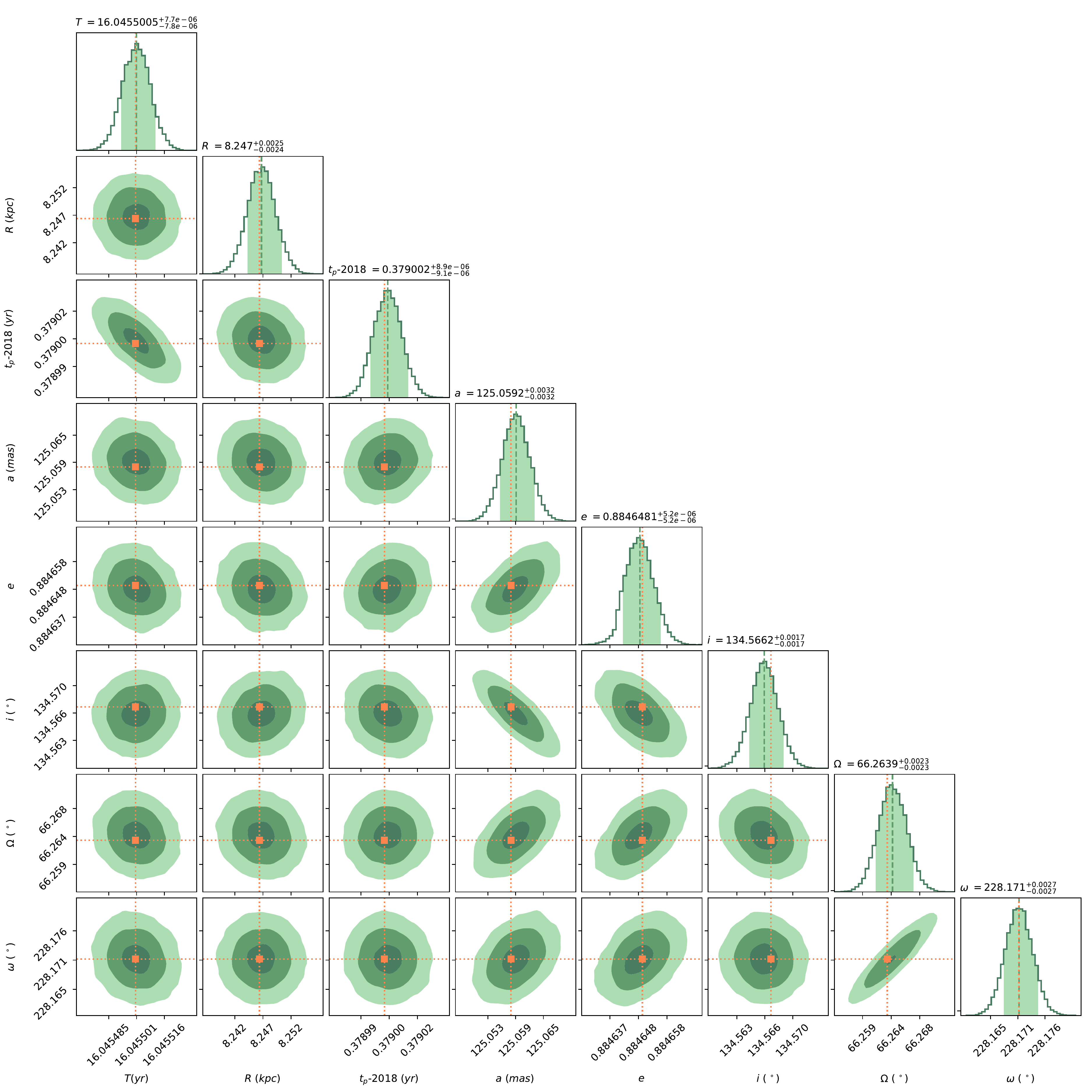}
    	\caption{The Figure depicts the posterior distribution of the 8-dimensional parameter space of our model obtained by using mock observations and fixing the geometry of space-time to a Schwarzschild BH. Orange squares and dotted lines report the \emph{true} values, i.e. the ones used to build the mock catalogue itself and reported in Table \ref{tab:s2_gravity}.}
    	\label{fig:posterior_mock_data_sch}
    \end{figure}

    We use uniform priors on the orbital parameter 
    centered on the best fitting values from \cite{gravity} with an amplitude that is 10 times the experimental error reported in Table \ref{tab:s2_gravity}. In Figure \ref{fig:posterior_mock_data_sch} we report the 1$\sigma$, 2$\sigma$  and 3$\sigma$ contours of the posterior distribution obtained from this analysis. As expected, the contours are centered on our input (\emph{true}) values depicted as filled  orange squares and illustrating that they are well within the 1$\sigma$ interval of our posterior distribution. This result validates the reliability of our mock catalogue for S0-2 and, more in general, the effectiveness of our fitting procedure.
    
    \subsection{Results}    \label{sec:mock_results}

\begin{figure}[htbp]
    	\includegraphics[width = \textwidth]{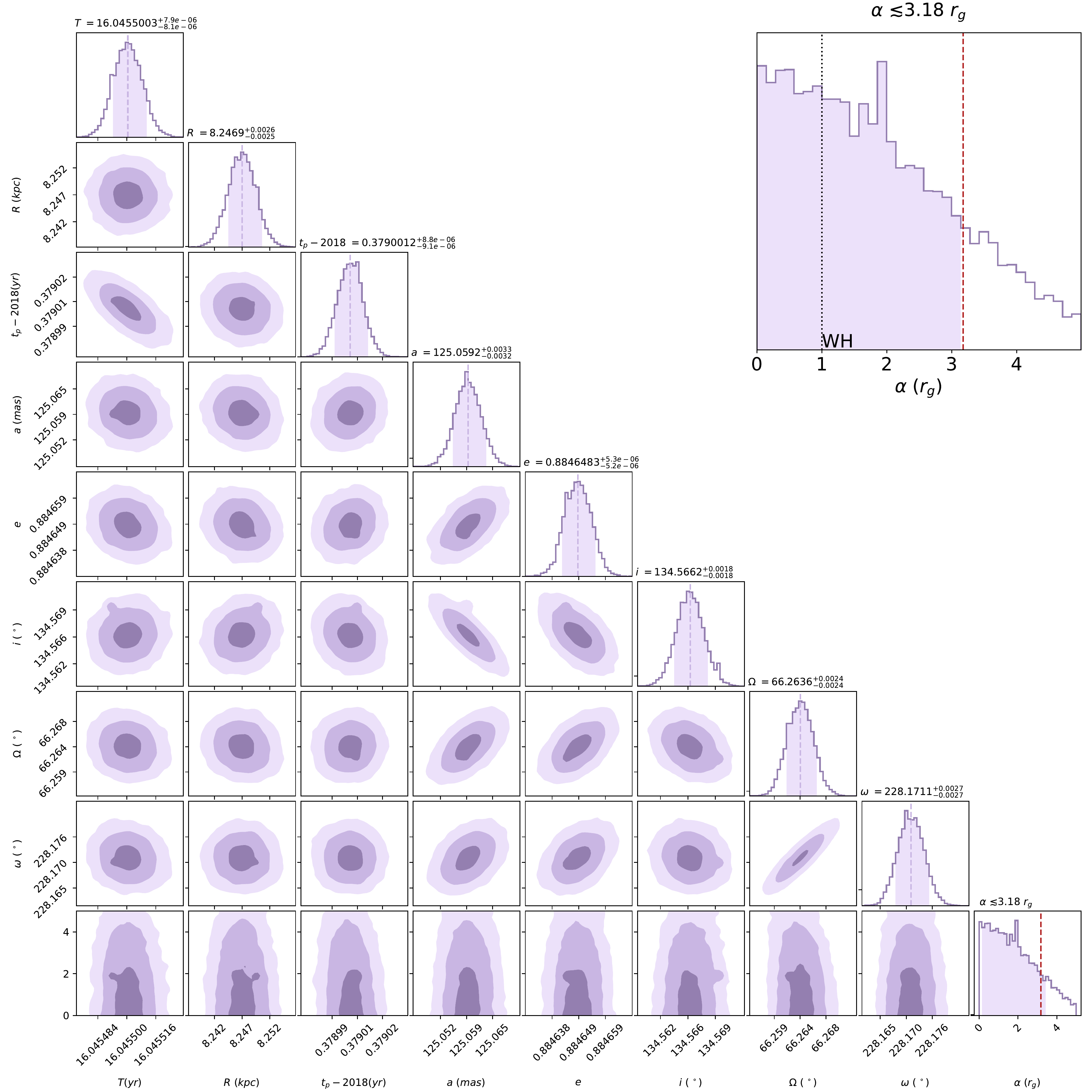}
    	\caption{The figure illustrates the  posterior distribution of the 9-dimensional parameter space of our model when using mock observations. At the present stage of precision we are not able to distinguish between a BH and and WH nature (the limit is marked by a dotted vertical black line).}
    	\label{fig:posterior_mock_data}
    \end{figure}

    Finally, we run an additional MCMC analysis leaving all the parameters free to vary, including the parameter $\alpha$ in Eq. \eqref{eq:metric}. Since the mock catalogue has been generated by assuming a Schwarzschild BH  nature for SgrA*, our reference value for this parameter is $\alpha = 0$. Thus, from the posterior distribution of the parameter $\alpha$ resulting from our analysis we can establish the ability of the upcoming observations with the GRAVITY/SINFONI instruments to rule out the WH nature for SgrA*. In particular, in Figure \ref{fig:posterior_mock_data} we report the full posterior distribution of the 9-dimensional parameter space of this analysis, highlighting its 1$\sigma$, 2$\sigma$  and 3$\sigma$ contours and the posterior of each parameter. As in Figure \ref{fig:posterior_with_data}, the inset plot reports the histogram for the parameter $\alpha$. The shaded area corresponds to the region within the 84-percentile of the distribution, the vertical dashed red line, thus, marks its inferred upper limit, while the black vertical line denotes the one-way transversable WH, separating the two classes of natures for SgrA*. As it appears from our analysis, reaching an astrometric precision as that nominally provided by the GRAVITY instrument, over two entire orbits of the star, is not yet sufficient to distinguish between the BH nature and the WH nature of the massive object in the GC, being able to only provide with an upper limit $\alpha \lesssim 5 r_G$ thus excluding WHs with a larger throat at $3\sigma$. Nevertheless, this would represent an improvement of a factor $\sim 25$ with respect to the current data.

    \section{The role of hypothetical closer stars} \label{sec:S62_toy}
    
    In sections \ref{sec:s2_data} and \ref{sec:s2_mock} our analysis focused  only on the S0-2 star. This is because, up to the date, it is the best studied star in the nuclear star cluster of Milky Way both in terms of observational accuracy and in terms of detectability of relativistic effects on its orbit. Nevertheless, we have demonstrated that neither publicly available data for S0-2 nor future observations with the most precise instrument available today for astrometry, GRAVITY, can effectively constrain the parameter space of a bouncing geometry as that in Eq. \eqref{eq:metric}, discerning the BH nature to that of a WH. However, stars orbiting closer to SgrA* than S0-2 have been hypothesized \cite{waisberg2018} to be able to investigate more accurately the gravitational field generated by it.
    For this reason, in section \ref{sec:toy_models} we have considered several toy models, whose orbital parameters have been chosen in order to have an increasingly closer pericentre passage and, hence, an increasingly higher departure between the orbits around a BH and a WH. Finally, we have picked one of the toy models and repeated our MCMC analysis showing at which extent GRAVITY-like observations of such hypothetical object could exclude the WH nature of the central object, depending on the observational strategy.
    
    \subsection{Using toy models to measure the Black Bounce metric}
    \label{sec:toy_models}
    
\begin{figure}[ht]
        \centering
        \includegraphics[width = \textwidth]{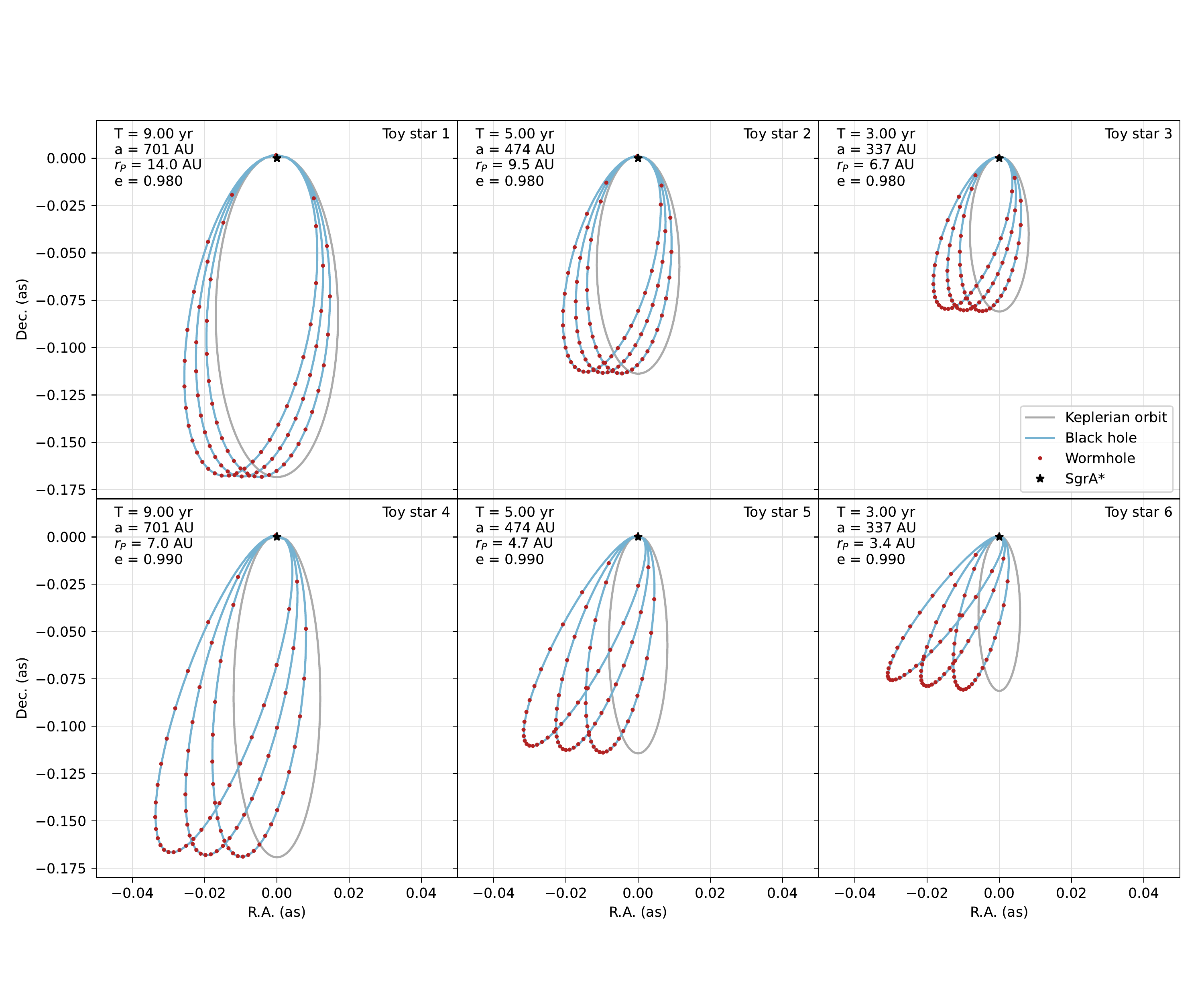}
        \caption{The figure depicts the orbital motion of the toy models listed in Table \ref{tab:toy_models}. The gray lines report the corresponding Keplerian orbit. 
        Geodesic trajectories are reported for a Schwarzschild BH ($\alpha = 0$, turquoise line) and for a one-way transversable WH ($\alpha = r_g$, red dots).}
        \label{fig:toy_models_orbits}
    \end{figure}

  Here we present a series of toy models describing hypothetical stars in the GC orbiting SgrA* closer than S0-2. 
  Our aim is to investigate whether such an object may serve to shed light on the fundamental nature of the central supermassive object whose mass is set to $M = 4.216\times10^6M_\odot$ and the distance of the GC to $R = 8.246$ kpc for all toy models. The orbital parameters of those synthetic stars have been chosen in order to probe increasingly smaller distances at pericentre, with the idea that the closer the star gets to SgrA*, the more the orbital motion becomes sensible to the geometry of space-time. In particular, we have chosen three orbital periods of nine, five and three years respectively and for each of them we have fixed the orbital eccentricity to either 0.98 or 0.99. Kepler's third law fixes the semi-major axis through the relation: $a = \sqrt[3]{GMT^2/4\pi^2}$; and the distance of the star at pericentre can be estimated as $r_p = a(1-e)$. We set for convenience 2025.00 as time of next pericentre passage for all the stars, without loss of generality\footnote{Since the metric coefficient in Eq. \eqref{eq:metric} do not explicitly depend on the coordinate time $t$, the problem of motion is symmetric with respect to time translations and hence, regardless of the particular value of $t_p$ we would draw the same conclusions.}.
    The three angular orbital parameters ($i$, $\Omega$, $\omega$) have been set to zero in order to guarantee that the orbit is always parallel to the sky-plane and that the astrometric shift of the star resulting from its periastron advance is more prominent in the R.A.  direction. 
    In other words, the orbits always start at pericentre with the semi-major axis that is in a vertical position with respect to the R.A. -Dec reference system. This means that, as the orbit's periastron advances, the shift is mostly noticeable in the R.A.  direction and we can focus on that particular direction, as it appears in Figure \ref{fig:toy_models_orbits}. 
    Moreover, since the orbit hasn't got any inclination with respect to the celestial plane of a distant observer, we do not  need to account for the Rømer delay. 
    \begin{table}[t]
        \centering
        \renewcommand{\arraystretch}{1.5}
        \setlength{\tabcolsep}{15pt}
        \begin{tabular}{lcccc}
            \hline
            Model & $T$  & $a$  & $r_p$  & $e$ \\ 
             & (yr) & (AU) & (AU) & \\\hline
            Toy star 1 & 9 & 701 & 14.0 & 0.98 \\
            Toy star 2 & 5 & 474 & 9.5 & 0.98 \\
            Toy star 3 & 3 & 337 & 6.7 & 0.98 \\
            Toy star 4 & 9 & 701 & 7.0 & 0.99 \\
            Toy star 5 & 5 & 474 & 4.7 & 0.99 \\
            Toy star 6 & 3 & 337 & 3.4 & 0.99 \\ \hline
        \end{tabular}
        \caption{The orbital parameters for toy models of stars orbiting SgrA*. Column 1 lists the name used to label the toy model; Columns two and three list the orbital period and the semi-major axis (related through the third Kepler's law); Column 4 lists the distance of the pericentre from SgrA*; finally, Column 5 lists the eccentricity.
        For all the toy models the three angular orbital parameters ($i$, $\Omega$, $\omega$) have been set to zero and, additionally, the mass and distance of SgrA* are fixed to the ones from \cite{gravity}.}
        \label{tab:toy_models}
    \end{table}
    Although we are going to use
    toy models,
    it is worth to note that a population of faint fast-moving stars in the GC \cite{Peisker2020, gravity2021} 
    could be 
    detected near in the future by means of NIR observations carried out with the Extremely Large Telescope (ELT) \cite{MICADO}.
    
    The numerically integrated, sky-projected orbits for these toy models are illustrated in Figure \ref{fig:toy_models_orbits} where we report both the orbit for the BH model (as a blue line) and the one for a WH with $\alpha = 1 r_g$ (as a red dotted line), alongside the corresponding Keplerian orbits (as a grey line). As it appears from the figure the effect of the periastron advance is strongly evident by the fact that the orbits do not appear to be closed (as their Keplerian counterparts), but at this scale no departure appears to be visible between the BH and WH solutions. 
        \begin{figure}[ht]
        \centering
        \includegraphics[width = \textwidth]{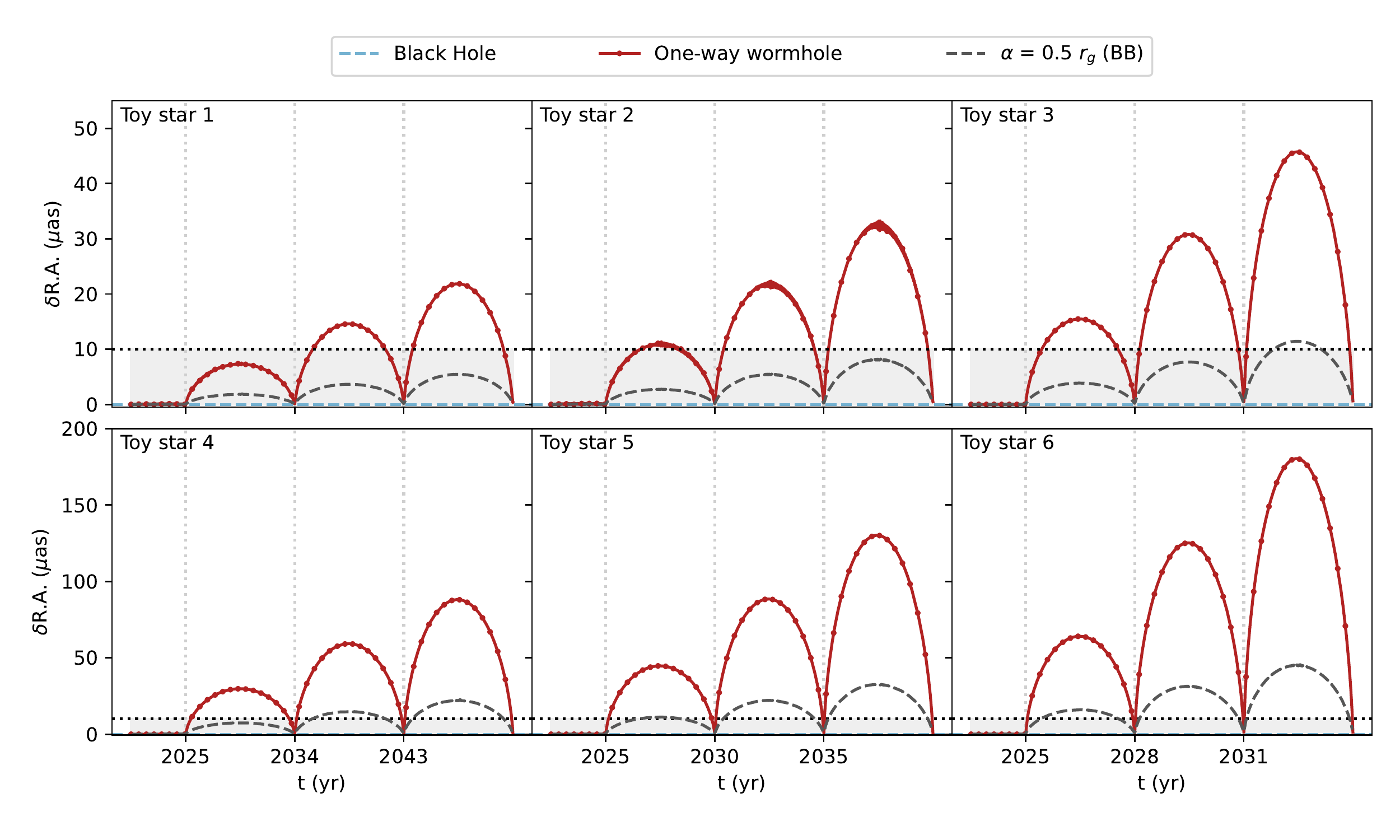}
        \caption{A focus on the departure between the geodesic orbit around a BH (which we take as a reference at level 0) and one way transversable WH (the red line) in the R.A. direction. The gray shaded area depicts the nominal astrometric accuracy of the GRAVITY interferometer, showing that observations with GRAVITY can actually be able to detect departures between the two geometries for the toy models we have considered, that is more noticeable as the orbital parameters gets more \emph{extreme}. Moreover, we report the geodesic orbit (dashed gray line) for a regular BH with $\alpha = 0.5\,r_g$ as an intermediate stage between the BH and WH solutions.}
        \label{fig:toy_models_ra}
    \end{figure}
    
    To make this departure more clear, in Figure \ref{fig:toy_models_ra} we focus only on the R.A.  direction, reporting both the WH solution (for $\alpha = 1r_g$) and an intermediate BB solution for $\alpha = r_g/2$ (corresponding to a regular BH) relative to the BH one, i.e. obtained subtracting the right ascension of the BH solution from that of these models, as a red solid and a black dashed line, respectively. The figure also reports as a shaded region the nominal astrometric sensitivity threshold of the GRAVITY interferometer. As it appears, the departure (with respect to a case where SgrA* is a Schwarzschild BH)  of the orbital motion of all our toy stars eventually (over multiple orbital periods, in some cases) gain enough astrometric shift to be, in principle, detected using observations. This means that such objects could actually serve as probes to rule out either the WH or BH nature of SgrA*. As expected, the greatest departure is observed at apocentre and, depending on the orbital parameters, this departure surpasses the detectability threshold already after one orbit (for Toy stars 3, 4, 5 and 6) or only after multiple orbits (Toy stars 1 and 2). Assuming to have observations that cover only one entire orbital period (from one apocentre to the next one), we focused our attention on only one of the toy models, namely Toy star 5, and we built different mock catalogues (following the same algorithm described in section \ref{sec:mock_catalogue}) investigating different observational strategies to assess how can the WH nature be ruled out (assuming that the true value is $\alpha = 0$). Since we have demonstrated that the greatest astrometric discrepancy between the orbits of the star around a BH and a WH is recorded at apocentre we postulate that observations should be especially dense around apocentre in order to capture the biggest deviation. This comes under the assumption that no extended mass distribution is present around SgrA* whose effect on the periastron advance would dominate over the general relativistic orbital precession in the apocentre region, as pointed out in \cite{Heissel2021}. Moreover, observations at apocentre should be complemented with observations at pericentre and in the rest of the orbit, in order to properly constrain the orbital period of the star, its eccentricity and its time of pericentre passage. The different observational strategies will be characterized by a different number of data points (and hence night of observations) in different regions of the orbit. In particular, we divided the orbit of the toy star in three regions. These are reported in Figure \ref{fig:orbit_regions} and correspond to:
    \begin{itemize}
        \item region $\mathcal{A}$ is centered on the star's pericentre and is given by the points along the orbit whose eccentric relativistic anomaly $\mathcal{E}$ is within 60$^\circ$ of the argument of pericentre. The relativistic anomaly is an angle that is useful to introduce in GR to account for the orbital precession of massive test particles \cite{Chandrasekhar1983}. The true anomaly of the star, which is the angular coordinate $\phi$ whose evolution is described by Eq. \eqref{eq:geodesic_phi}, varies between 0 and $2\pi$. Nevertheless, due to the periastron advance, apocentre and pericentre are not always represented by the same angular coordinates $\phi$, which evolve over time by the factor $\Delta\omega$ in Eq. \eqref{eq:gr_precession}, for the case of a Schwarzschild BH. The relativistic anomaly $\chi$, on the other hand, is defined by the relation
        \begin{equation}
            \frac{1}{r} = \frac{1+e\cos\chi}{\ell},
        \end{equation}
        where $\ell$ is the orbit's latus rectum, given by $\ell = a(1-e^2)$. By definition, the coordinate $r$ has always a maximum (apocentre) for $\chi = \pi$ and always a minimum (pericentre) in $\chi = 0$. This means that the angle $\chi$ describes positions on the geodesic, as the ellipse precedes. Finally, we have computed the eccentric relativistic anomaly $\mathcal{E}$ in order to change the origin of the angles from the central mass (located in one of the ellipse's foci) to the centre of the ellipse, which is required if we want to select angles symmetrically with respect to fixed point on the star's path. The relation between the anomaly and the eccentric anomaly is given by:
        \begin{equation}
            \tan\frac{\chi}{2} = \sqrt{\frac{1+e}{1-e}}\tan\frac{\mathcal{E}}{2}.
        \end{equation}
        In Figure \ref{fig:orbit_regions} we report a visual illustration of the different regions and of the angular coordinates that we have employed.
        \item region $\mathcal{B}$ is built with the same idea of region $\mathcal{A}$ but is centered on the star's apocentre ($\chi = \pi$), instead.
        \item region $\mathcal{C}$ covers the rest of the orbit.
    \end{itemize}

\begin{figure}[h]
    \centering
    \includegraphics[width = \textwidth]{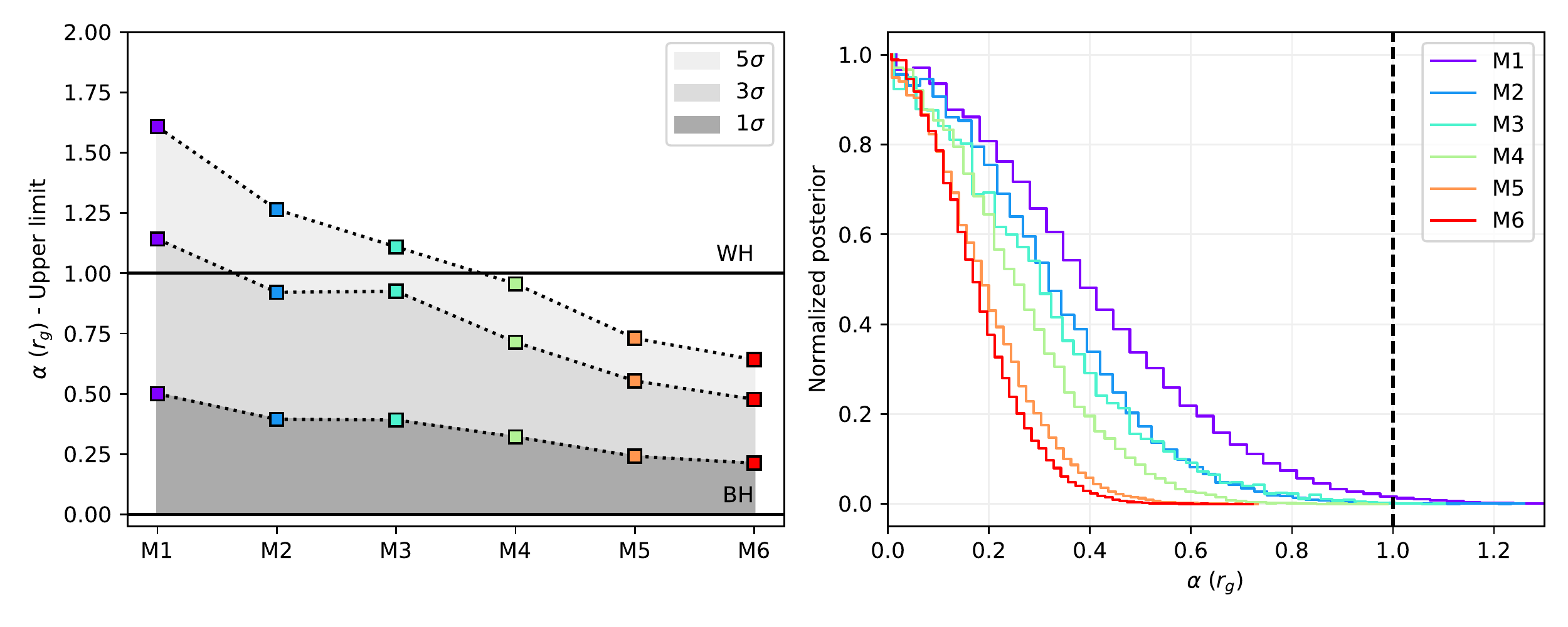}
    \caption{On the left panel we report the 1$\sigma$, 3$\sigma$ and 5$\sigma$ uppers bound on the parameter $\alpha$ resulting from the posterior distributions (reported on the right panel), as a function of the observational strategy for the Toy star 5 in Table \ref{tab:toy_models}.}
    \label{fig:mock-toy-alpha}
\end{figure}

    We have thus built six mock catalogues of observations (using the same procedure reported in section \ref{sec:mock_catalogue}) whose number of observations for each region are reported in Table \ref{tab:mock_catlogues}. The idea behind this choice of numbers is to assess to which extent the boundaries that we can put on the parameter $\alpha$ improve or get worst as the number of observational points in each region changes. The model M6 is our best case scenario and is characterized by the same number of observations per region as the mock catalogue that we have built for S0-2 in section \ref{sec:s2_mock}. All the other mock catalogues have a fewer number of observations in each region, reflecting the fact that such a close and fast-moving star as our Toy Model 5 could represent a greater observational challenge than S0-2. Mock catalogues M5 and M4, thus, have increasingly less points at pericentre, but maintaining the same number of observations in regions $\mathcal{B}$ and $\mathcal{C}$. In the mock catalogue M3 we have reduced by a half, with respect to M4, the number of observations in each region, leading to a total of 30 observations over an entire orbit. Since we have shown in Figure \ref{fig:toy_models_ra} that the maximum astrometric deviation between a BH and a WH model happens around apocentre, in the mock catalogue M2 we have maintained the total number of observations to 30, but we moved the observations in region $\mathcal{B}$ to region $\mathcal{C}$. Finally, in our most drastic scenario, M1, we have reduced by $\sim 1/3$ the total number of observations, leaving only 5 data points around pericentre and $6$ in the rest of the orbit, with no observations in region $\mathcal{B}$. For each of this mock catalogues we have performed a MCMC analysis similarly to what we have done in section \ref{sec:mock_results}. We have left all the 9 parameters ($R$, $T$,  $t_p$, $a$, $e$, $i$, $\Omega$, $\omega$, $\alpha$) free to vary and we have assigned uniform large priors on all the parameters, centered on the true value on which we have built the catalogue itself. These are reported in Table \ref{tab:priors_toy}, along with the best fitting values for the parameters, for the different mock catalogues. Moreover, in Figure \ref{fig:mock-toy-alpha} we report the posterior for the parameter $\alpha$ and the upper 1$\sigma$, 3$\sigma$ and 5$\sigma$ bounds that we have obtained for the different observational strategies. As expected, the constraining power on the parameter $\alpha$ of our datasets is increasingly higher, leading to a significant improvement on the upper bounds placed on the parameter $\alpha$ going from our worst to the best case scenario. In particular, the data in M1 are only able to exclude the WH nature of the central object at 1$\sigma$ significance , with the limit $\alpha = 1r_g$ well within both the 3$\sigma$ and 5$\sigma$ credible intervals. Both M2 and M3, were able to exclude the WH nature at 3$\sigma$ significance, but the 5$\sigma$ upper limit on $\alpha$ is above the $\alpha = 1r_g$ limit and slightly improves between M2 and M3, due to the additional information carried by observation recorded at apocentre. M4 is the first dataset that succeeds in excluding at more than 5$\sigma$ the WH nature of Sgr A*, with M5 and M6 increasingly improving this result. 

    \begin{table}[t]
        \renewcommand{\arraystretch}{1.5}
        \setlength{\tabcolsep}{12pt}
        \centering
        \begin{tabular}{lccc}\hline
                Mock catalogue & $\mathcal{A}$ & $\mathcal{B}$ & $\mathcal{C}$\\ \hline
                M1 & 5 & 0 & 6\\
                M2 & 10 & 0 & 20\\
                M3 & 10 & 10 & 10\\
                M4 & 20 & 20 & 20\\
                M5 & 35 & 20 & 20 \\
                M6 & 45 & 20 & 20\\
                 \hline
        \end{tabular}
        \caption{We report the number of observations that are performed in the regions of the toy star's orbit that we have defined in Figure \ref{fig:orbit_regions} for the six different mock catalogues that we have built.}
        \label{tab:mock_catlogues}
    \end{table}
        
    \noindent
    \begin{minipage}[t]{\textwidth}
        \vspace{1cm}
            \begin{minipage}[t!]{0.48\textwidth}
                \centering
                \includegraphics[width = 0.78\textwidth]{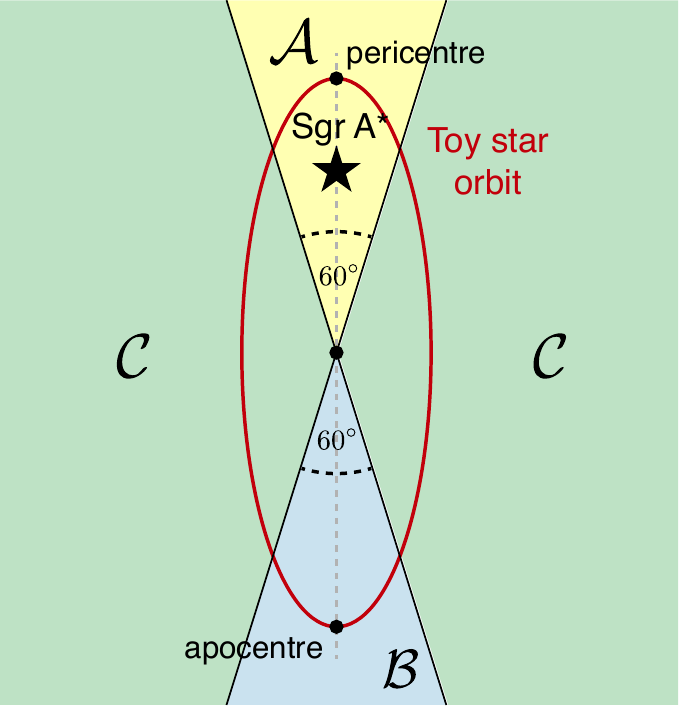}
            \end{minipage}
            \hfill
            \begin{minipage}[t!]{0.5\textwidth}
                \centering
                \includegraphics[width = \textwidth]{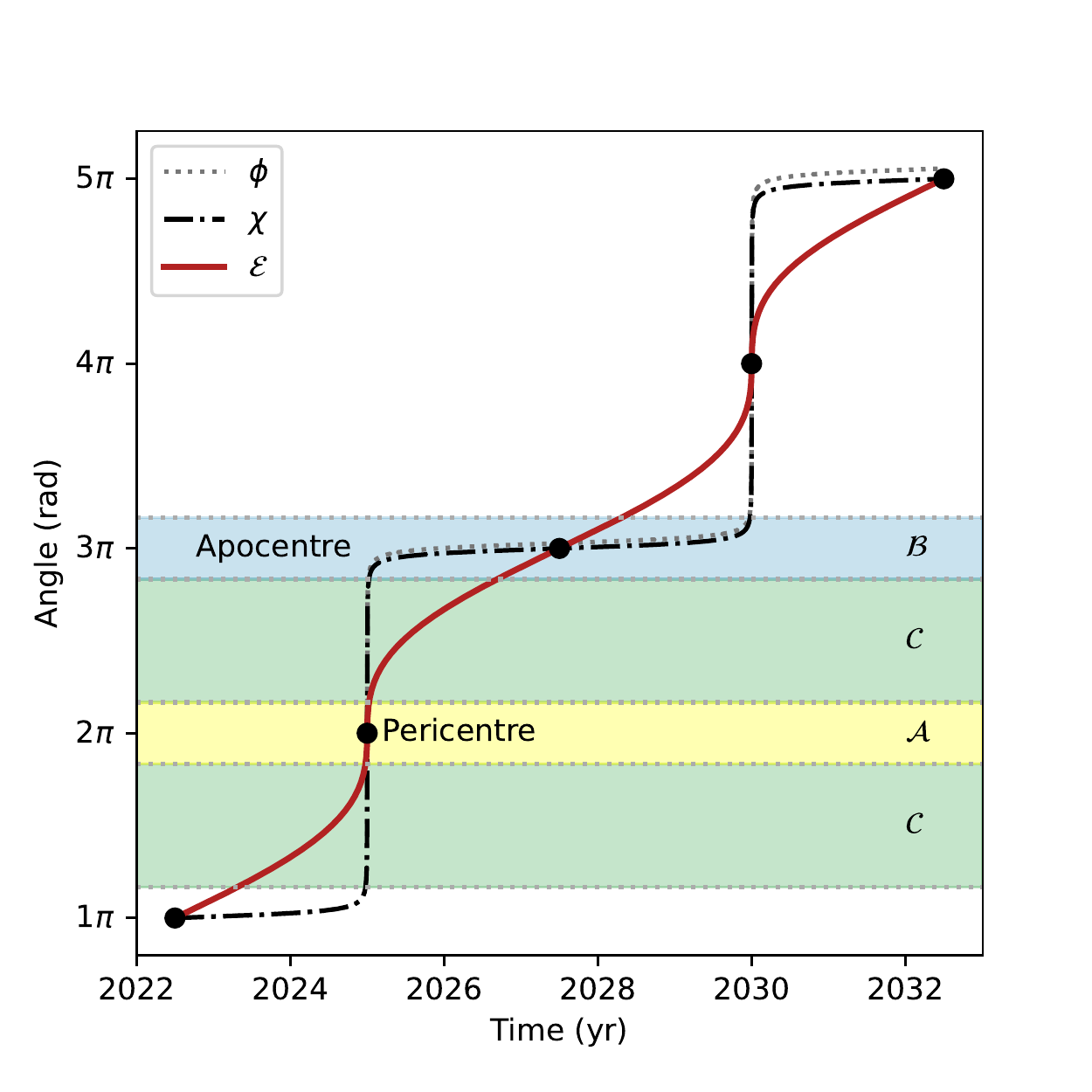}
            \end{minipage}
            \captionof{figure}{\emph{Left panel:} We have divided the orbit into three regions: region $\mathcal{A}$ (in yellow) is centered on the pericentre of the orbit and has an (eccentric anomaly) amplitude of 60$^\circ$; region $\mathcal{B}$ (light blue) has the same amplitude of $\mathcal{A}$ but is centered on the star's apocentre; region $\mathcal{C}$ (in green) covers the rest of the orbit. The red line must be imagined as an ellipse following the star during the periastron advance and the angles representing the relativistic eccentric anomaly $\mathcal{E}$. \emph{Right panel:} the actual values of the different angles along the integrated orbit. The red line represents the relativistic eccentric anomaly $\mathcal{E}$, the black dash-dotted line the relativistic anomaly $\chi$, while the gray dotted line is the true anomaly $\phi$ of the star (which gradually piles up a shift with respect to $\chi$, due to the orbital precession).}
            \label{fig:orbit_regions}
    \end{minipage}

    \begin{table}[h]
        \centering
        \renewcommand{\arraystretch}{1.5}
        \setlength{\tabcolsep}{5pt}
        \resizebox{\textwidth}{!}{%
        \begin{tabular}{lcccccc}
            \hline
            & & \multicolumn{2}{c}{Uniform prior} & & &\\ \cline{3-4}
            Parameter & True & Start & End & \textbf{M1} & \textbf{M2} & \textbf{M3} \\ \hline
            $R$ (kpc) & 8.246 & 7 & 9 & $8.249\pm0.023$ & $8.246\pm0.012$ & $8.248\pm0.011$ \\
            $T$ (yr) & 5 & 4 & 6 & $5.0007\pm0.0062$ & $5.0000\pm0.0023$ & $5.00004\pm0.00049$ \\
            $t_p-2025$ (yr) & 0 & 2024 & 2026 & $0.0000008\pm0.0000076$ & $0.0000004\pm0.0000056$ & $0.0000004\pm0.0000057$ \\
            $a$ (AU) & 474 & 450 & 500 & $474.1\pm1.4$ & $474.12\pm0.72$ & $474.06\pm0.66$ \\
            $e$ & 0.99 & 0.98 & 1 & $0.990002\pm0.000035$ & $0.990001\pm0.000019$ & $0.990002\pm0.000017$ \\
            $i$ ($^\circ$) & 0 & -10 & 10 & $0.06\pm0.33$ & $0.02\pm0.21$ & $0.01\pm0.20$ \\
            $\omega$ ($^\circ$) & 0 & -10 & 10 & $0.10\pm0.56$ & $0.04\pm0.45$ & $0.06\pm0.59$ \\
            $\Omega$ ($^\circ$) & 0 & -10 & 10 & $0.08\pm0.58$ & $0.08\pm0.46$ & $0.08\pm0.61$ \\
            $\alpha$ ($r_g$) & 0 & 0 & 10 & $\lesssim1.61\, (5\sigma)$ & $\lesssim1.26\, (5\sigma)$ & $\lesssim1.11\, (5\sigma)$ \\
            \hline
             & & & & \textbf{M4} & \textbf{M5} & \textbf{M6} \\ \hline
            $R$ (kpc) & 8.246 & 7 & 9 & $8.2467\pm0.0053$ & $8.2469\pm0.0049$ & $8.2466\pm0.0049$ \\
            $T$ (yr) & 5 & 4 & 6 & $5.00004\pm0.00034$ & $5.00000\pm0.00032$ & $5.00005\pm0.00032$ \\
            $t_p-2025$ (yr) & 0 & 2024 & 2026 & $0.0000007\pm0.0000039$ & $0.0000003\pm0.0000030$ & $0.0000001\pm0.0000027$ \\
            $a$ (AU) & 474 & 450 & 500 & $474.03\pm0.31$ & $474.01\pm0.29$ & $474.01\pm0.28$ \\
            $e$ & 0.99 & 0.98 & 1 & $0.9900001\pm0.0000064$ & $0.9900003\pm0.0000061$ & $0.9900008\pm0.9900060$ \\
            $i$ ($^\circ$) & 0 & -10 & 10 & $0.0011\pm0.0095$ & $0.0006\pm0.0048$ & $0.0007\pm0.0038$ \\
            $\omega$ ($^\circ$) & 0 & -10 & 10 & $0.08\pm0.41$ & $0.06\pm0.35$ & $0.01\pm0.35$ \\
            $\Omega$ ($^\circ$) & 0 & -10 & 10 & $0.06\pm0.42$ & $0.04\pm0.36$ & $0.05\pm0.35$ \\
            $\alpha$ ($r_g$) & 0 & 0 & 10 & $\lesssim0.96\, (5\sigma)$ & $\lesssim0.73\, (5\sigma)$ & $\lesssim0.64\, (5\sigma)$ \\
            \hline
        \end{tabular}}
        \caption{A summary of the priors and best-fit values from our MCMC analysis on the different mock catalogues that we have built for the Toy star 5. More specifically, column 2 reports the true value of the fit parameters, i.e. the one used to build the mock catalogue; columns 3 and 4 report the upper and lower limits of the uniform distributions that we used as priors; column 5 to 7 report the best fitting values resulting from the MCMC posterior analysis with their 1$\sigma$ credible intervals and $5\sigma$ upper confidence limit for the parameter $\alpha$, also reported in Figure \ref{fig:mock-toy-alpha}.}
        \label{tab:priors_toy}
    \end{table}

\section{Conclusions and Discussions}\label{sec:conclusions}

The nuclear star cluster of the S-stars, orbiting the compact supermassive radio source SgrA* that lies in the centre of the Milky Way, provides with a formidable laboratory to test the theory of gravity. While observations in the last thirty years have provided with compelling evidences the SgrA* is indeed a SMBH as described by GR, its intrinsic nature is still object of debate. A viable alternative to the BH model is that SgrA* has a WH nature. To investigate the feasibility of such a BH mimicker, different studies have analyzed the observable signatures arising from this kind of geometries. An example is the appearance of additional light rings in the shadow cast by such an object upon the light emitted by its accretion flow \cite{Guerrero2021, 2021arXiv211010002O}.
Also, in \cite{Cheng2021}, authors have predicted the accuracy down to which departures from a static BH model can be detected, by mirroring the astrometric accuracy of the instrument GRAVITY.

In this work we have studied the uniparametric family of BB solutions (firstly introduced in \cite{simpson2019}) that continuously map between a Schwarzschild BH model and a two-way transversable WH. In particular, exploiting current observations of the S0-2 stars, and forecasting future observations of the same object by  GRAVITY, we have investigated the current and future possibility to constrain the departure from a Schwarzschild BH. To this aim, we have used publicly available data for the orbital motion of the S0-2 star and we have estimated the departure from a Schwarzschild BH using an MCMC algorithm. Our analysis have demonstrated that current observations do not yet provide with sufficient constraining power to exclude neither the BH nor the WH nature. Hence, we proceeded further with our analysis and investigated whether an improved astrometric dataset of the same object could be able to rule-out one of the two models. This led us to the development of a mock catalogue for S0-2 that would mimic future observations performed with the GRAVITY instrument, both in terms of precise astrometry and of observational strategy. Firstly, we have assessed the reliability of our mock catalogue and of our fitting procedure, in general, demonstrating its effectiveness in recovering our baseline model (see Figure \ref{fig:posterior_mock_data_sch}). Then, we have applied the fitting algorithm on the mock catalogue by leaving the parameter $\alpha$ free. Since the mock catalogue was generated on an $\alpha = 0$ model, we were able to test the ability to rule out the WH nature with future observations. The results of our analysis, shown in Figure \ref{fig:posterior_mock_data}, demonstrate that even with a much more precise astrometric dataset we would not be able to unambiguously discern between the BH and WH natures for SgrA*, using the orbit of the S0-2 star. 

In order to investigate further the ability to constrain the WH geometry of space-time around SgrA* with the orbits of the S-stars, we studied the motion of closer stars for which the effects related to the modification of the geometry are expected to be larger.  
For this reason, we built six toy models, whose orbital parameters are reported in Table \ref{tab:toy_models}, which investigate increasingly more extreme regions of the parameter space. The orbits of these toy models and the corresponding discrepancies between the BH and WH models are reported in Figures \ref{fig:toy_models_orbits} and \ref{fig:toy_models_ra}, respectively. As we have demonstrated, objects with such orbital features can effectively be used as probes to detect departures from the BH geometry within the current sensitivity of astrometric instruments as GRAVITY. Depending on the proximity reached during pericentre, the astrometric deviation surpasses the sensitivity threshold of $\sim10\,\mu$as within either one or multiple orbital periods, as reported in Figure \ref{fig:toy_models_ra}. In order to assess quantitatively the ability to constraint the BB geometry using observations of such hypothetical stars, we have built mock catalogues for one of these toy models reflecting different observational strategies (reported in Table \ref{tab:mock_catlogues}). By repeating the MCMC analysis on each of these mock catalogues we analyzed the posterior distribution on the parameter $\alpha$ and its $1\sigma$, $3\sigma$ and $5\sigma$ confidence upper limits (all reported in Figure \ref{fig:mock-toy-alpha}). Our analysis suggests that in order to distinguish at a 5$\sigma$ significance level the BH nature to that of a WH for SgrA*, a uniform astrometric coverage of at least 20 observations per orbital region is required for a 5-year-period highly-eccentric (0.99) star in the GC, maintaining a nominal astrometric accuracy of $10\,\mu$as over the entire observational campaign.
Finally, it is worth to note that rotation of the central supermassive object is not taken into account in this work due to the lack of direct observational probes of its rotation. Nevertheless, assessing the impact of the rotation on our analysis will be the subject of forthcoming studies.

\acknowledgments

We are grateful to the referee for his/her valuable comments.
RDM acknowledges support from Consejeria de Educación de la Junta de Castilla y León and from the Fondo Social Europeo.
IDM acknowledges support from Ayuda  IJCI2018-036198-I  funded by  MCIN/AEI/  10.13039/501100011033  and:  FSE  “El FSE  invierte  en  tu  futuro”  o  “Financiado  por  la  Unión  Europea   “NextGenerationEU”/PRTR. 
IDM is also supported by the project PGC2018-096038-B-I00  funded by the Spanish "Ministerio de Ciencia e Innovación" and FEDER “A way of making Europe", and by the project SA096P20 Junta de Castilla y León.

\bibliographystyle{JHEP}
\bibliography{biblio}

\end{document}